# Ground-State Energy as a Simple Sum of Orbital Energies in Kohn-Sham Theory:

# A Shift in Perspective through a Shift in Potential

*Mel Levy[a,b,c,1] and Federico Zahariev[d,2]*

a  Department of Chemistry, Duke University, Durham, North Carolina 27708

b  Department of Physics, North Carolina A&T State University, Greensboro, North Carolina 27411

c  Department of Chemistry and Quantum Theory Group, Tulane University, New Orleans, Louisiana 70118, USA

d  Department of Chemistry and Ames Laboratory, Iowa State University, Ames, Iowa 50011, USA

(September 2, 2014)

It is observed that the exact interacting ground-state electronic energy of interest may be obtained directly, in principle, as a simple sum of orbital energies when a universal density-dependent term is added to $w([\rho];r)$, the familiar Hartree plus exchange-correlation component in the Kohn-Sham effective potential. The resultant shifted potential, $\bar{w}([\rho];r)$, actually changes less on average than $w([\rho];r)$ when the density changes, including the fact that $\bar{w}([\rho];r)$ does not undergo a discontinuity when the number of electrons increases through an integer. Thus the approximation of $\bar{w}([\rho];r)$ represents an alternative direct approach for the approximation of the ground-state energy and density.

[1] mlevy@tulane.edu

[2] fzahari@iastate.edu

Exact solutions of the Schrödinger wavefunction equation yield the necessary information for ascertaining the electronic structures of atoms, molecules and solids. But these solutions are only obtainable for very simple systems. With this in mind, Hohenberg-Kohn-Sham density functional theory (DFT) was developed as a viable alternative to the use of the Schrödinger equation. While the dimension of the wavefunction grows with increasing electron number, the electron density contains only three dimensions, independent of the size of the system of interest.

Indeed, DFT now provides the most popular method for obtaining approximations to the ground-state energy $E_{GS}$ and ground-state density, $\rho_{GS}(r)$, of

$$\hat{H} = \sum_{i=1}^{N} v(r_i) + \hat{T} + \hat{V}_{ee}, \tag{1}$$

where $v(r)$ is the multiplicative external (usually electron-nuclear attraction) potential of interest, $\hat{T}$ is the kinetic energy operator, and $\hat{V}_{ee}$ is the electron-electron repulsion operator.

In DFT,

$$E_{GS} = \min_{\rho}\left\{T_s[\rho] + \int v(r)\rho(r)dr + G[\rho]\right\}, \tag{2}$$

where $\rho(r)$ is a trial electron density, $T_s[\rho]$ is its known noninteracting Kohn-Sham (KS) [1] kinetic energy, and $G[\rho]$ is the unknown universal functional of $\rho(r)$ that must be approximated. Specifically, $G[\rho]$ is the Hartree energy plus the exchange-correlation energy.

In traditional KS-DFT, $\rho_{GS}(r) = \sum_{i=1}^{N}|\varphi_i^{GS}(r)|^2$, where the $\varphi_i^{GS}(r)$ are obtained from the following converged non-interacting KS system [1], which results from the minimization in Eq. (2),

$$\left[-\frac{1}{2}\nabla^2 + v(r) + w([\rho_{GS}];r)\right]\varphi_i^{GS}(r) = \varepsilon_i \varphi_i^{GS}(r), \quad (i=1,2,...,N), \tag{3}$$

with $w([\rho_{GS}];r) = \left(\frac{\delta G[\rho]}{\delta \rho(r)}\right)\Big|_{\rho=\rho_{GS}}$. The $E_{GS}$ in Eq. (2) is then evaluated through

$$E_{GS} = \sum_{i=1}^{N} \langle \varphi_i^{GS} | -\frac{1}{2}\nabla^2 | \varphi_i^{GS} \rangle + \int v(r)\rho_{GS}(r)dr + G[\rho_{GS}]. \tag{4}$$

Thus the current procedure is to first obtain, in principle, the $\varphi_i^{GS}(r)$ and $\rho_{GS}(r)$ from the single-particle Kohn-Sham Euler equations, Eqs. (3), that arises from the minimization in Eq. (2). Then, the $\varphi_i^{GS}(r)$ and $\rho_{GS}(r)$ are inserted into Eq. (2) to obtain $E_{GS}$. Here we introduce a procedure where $E_{GS}$ is obtained, in principle, as the sum of one-electron orbital energies from a set of KS equations that has been modified by the presence of a density-dependent additive constant. Now, $E_{GS}$ is obtained directly from the same set of Euler equations that generate $\rho_{GS}(r)$ through a "direct-energy" KS theory.

To present our approach, we now define the potential $\bar{w}([\rho];r)$ as

$$\bar{w}([\rho];r) = w([\rho];r) + c[\rho] = w([\rho];r) + \frac{G[\rho] - \int w([\rho];r)\rho(r)dr}{\int \rho(r)dr}, \tag{5}$$

and observe that the $\varphi_i^{GS}(r)$ of Eq. (3) also satisfy

$$\left[-\frac{1}{2}\nabla^2 + v(r) + \bar{w}([\rho_{GS}];r)\right]\varphi_i^{GS}(r) = \bar{\varepsilon}_i \varphi_i^{GS}(r) \ (i=1,2,...,N), \tag{6}$$

because $\bar{w}([\rho];r)$ differs from $w([\rho];r)$ by only the density-dependent additive constant $c[\rho]$, which is defined in Eq. (5).

With $\bar{w}([\rho];r)$ defined by Eq. (5), it follows through multiplying it by $\rho(r)$ and integrating that

$$G[\rho] = \int \bar{w}([\rho];r)\rho(r)dr. \tag{7}$$

Thus by multiplying Eq. (6) by $\varphi_i^{GS}(r)$, summing over $i$, and then integrating, it should be clear by comparison with Eq. (4) that now $E_{GS}$ is the sum of orbital energies. Namely,

$$E_{GS} = \sum_{i=1}^{N} \bar{\varepsilon}_i, \qquad (8)$$

so that now both $\rho_{GS}(r)$ and $E_{GS}$ are simply obtained from the same system of equations, the new noninteracting system Eq. (6). [Alternatively, an orbital-free counterpart to Eq. (8) arises by the combination of Eq. (7) and (4) and with the kinetic-energy expression in Eq. (4) replaced by an orbital-free one. See [2] for a recent review of orbital-free kinetic-energies.]

$G[\rho]$ must be approximated. This is most commonly done without first approximating its functional derivative. Alternatively, it is well-known that $G[\rho]$ may be approximated by approximating its functional derivative along a designated path of densities and then performing a line integration [3,4]. Our approach introduces a third way to approximate $G[\rho]$, where it is simply obtained by integrating the product of $\bar{w}([\rho];r)$ with $\rho(r)$, Eq. (7). No line integrals are taken. Observe that $\bar{w}([\rho];r)$ serves both as a potential and as a local energy density per particle number [5]. Effectively, $G[\rho]$ is approximated by approximating $\bar{w}([\rho];r)$ directly in Eq. (7).

For illustrative purposes only, a very rough approximation to $\bar{w}([\rho];r)$ is

$$\bar{w}([\rho];r) \approx c_1 \int \frac{\rho(r')}{|r-r'|} dr' + c_2 \rho^{\frac{1}{3}}(r) + c_3 f\left(\frac{|\nabla \rho(r)|^m}{\rho(r)^n}\right), \qquad (9)$$

where the first two terms are suggested by the familiar local-density approximation for exchange. In Eq. (9), $c_1$, $c_2$, $c_3$, m, and n are parameters, and m, n, and the function $f(x)$ in the last term are such that $\lim_{|r|\to\infty} f\left(\frac{|\nabla \rho(r)|^m}{\rho(r)^n}\right)$ is non-zero, a property of the exact $\bar{w}([\rho];r)$.

Since the potential $\bar{w}([\rho];r)$ must be approximated, the objective is to have an approximate $\bar{w}([\rho];r)$ that satisfies as many known constraints as possible. By taking the functional derivative with respect to the density on both sides of Eq. (7), and by using Eq. (5) with the fact that $w([\rho];r) = \frac{\delta G[\rho]}{\delta \rho(r)}$, we obtain

$$\int \frac{\delta \bar{w}([\rho];r')}{\delta \rho(r)} \rho(r') dr' = -c[\rho] = -\lim_{|r| \to \infty} \bar{w}([\rho];r), \qquad (10)$$

for all $r$ on the left-hand side, which is a key constraint on $\bar{w}([\rho];r)$ for the correct approximation of it. We have also used the fact [6] that $\lim_{|r| \to \infty} w([\rho];r) = 0$. (For an interesting example of an approximate density-functional potential constructed in a different context but also having a density-dependent asymptotic value, see Ref [7].)

The satisfaction of Eq. (10) clearly provides the necessary and sufficient conditions that

$$\frac{\delta \int \bar{w}([\rho];r') \rho(r') dr'}{\delta \rho(r)} = w([\rho];r), \qquad (11)$$

which is a required constraint for $\bar{w}([\rho];r)$ to generate the exact KS orbitals, density, and ground-state energy.

To derive an intriguing inequality, let $\lambda$ signify any isoelectronic change in density, with $\rho_\lambda(r) = \rho(r)$ at $\lambda = 1$. [Among countless possibilities, an example is the coordinate scaled density $\rho_\lambda(r) = \lambda^3 \rho(\lambda r)$.] For an infinitesimal isoelectronic change in the density, the inequality results from the connection between $\frac{\partial \bar{w}([\rho_\lambda];r)}{\partial \lambda}$ and $\frac{\partial w([\rho_\lambda];r)}{\partial \lambda}$. To arrive at this inequality, observe that Eq. (5) and the fact that $w([\rho];r) = \frac{\delta G[\rho]}{\delta \rho(r)}$, lead to

$$\frac{\partial \bar{w}([\rho_\lambda];r)}{\partial \lambda} = \frac{\partial w([\rho_\lambda];r)}{\partial \lambda} - \int \frac{\partial w([\rho_\lambda];r)}{\partial \lambda} \bar{\rho}(r) dr, \qquad (12)$$

where $\bar{\rho}(r)$ is the density per particle, which is $\frac{\rho(r)}{\int \rho(r) dr}$.

Next square both sides of the above relation, multiply by $\bar{\rho}(r)$, and integrate to obtain the equality that relates $\frac{\partial \bar{w}([\rho_\lambda];r)}{\partial \lambda}$ to the "square of the uncertainty" in $\frac{\partial w([\rho_\lambda];r)}{\partial \lambda}$.

Namely,

$$\int \left(\frac{\partial \bar{w}([\rho_\lambda];r)}{\partial \lambda}\right)^2 \bar{\rho}(r) dr = \int \left(\frac{\partial w([\rho_\lambda];r)}{\partial \lambda}\right)^2 \bar{\rho}(r) dr - \left[\int \frac{\partial w([\rho_\lambda];r)}{\partial \lambda} \bar{\rho}(r) dr\right]^2, \quad (13)$$

from which it is clear that

$$\int \left(\frac{\partial \bar{w}([\rho_\lambda];r)}{\partial \lambda}\right)^2 \bar{\rho}(r) dr \leq \int \left(\frac{\partial w([\rho_\lambda];r)}{\partial \lambda}\right)^2 \bar{\rho}(r) dr. \quad (14)$$

Thus, upon any isoelectronic change in the density, inequality (14) reveals that $\bar{w}([\rho];r)$ actually changes less, on average, than $w([\rho];r)$. In fact, it can be shown that $\bar{w}([\rho];r)$ also changes less than any other potential that differs from $w([\rho];r)$ by a density-dependent additive constant. This is a tantalizing result.

We now discuss nonisoelectronic changes. For this purpose, note that any constant is annihilated when added to $w([\rho];r)$ in Eq. (5). This fact is important here for understanding the differences between the behaviors of $\bar{w}([\rho];r)$ and $w([\rho];r)$ during a fractional change [6,8] in electron number.

Because of its exchange-correlation component, $w([\rho];r)$ experiences a discontinuity [6] when the number of electrons is increased from $N-\delta$ to $N+\delta$, where $\delta$ is positive and infinitesimal. Based on theoretical arguments and studies of model systems [6,9], the potential $w([\rho_{N+\delta}];r)$ is shifted essentially by a constant with respect to $w([\rho_{N-\delta}];r)$ until approximately a cutoff distance $R(\delta)$ that depends upon $\delta$ and is far out in the region of very low density, and $w([\rho_{N+\delta}];r)$ is about the same as $w([\rho_{N-\delta}];r)$ beyond $R(\delta)$, where $\rho_M(r)$ is the ground-level density of M electrons. Further, $R(\delta) \to \infty$ in the limit as $\delta \to +0$, so that $\lim_{\delta \to +0}\left[w([\rho_{N+\delta}];r) - w([\rho_{N-\delta}];r)\right] = \Delta_{xc}$, for all $r$, where $\Delta_{xc}$ is a constant.

In contrast, it should be clear that $\lim_{\delta\to+0}\left[\bar{w}([\rho_{N+\delta}];r)-\bar{w}([\rho_{N-\delta}];r)\right]=0$, for all $r$. This is because $G[\rho]$ and the norm of $\rho(r)$ change continuously when the number of electrons increases from $N-\delta$ to $N+\delta$, and because a constant cancels out when it is added to $w([\rho];r)$ in Eq (5). Consequently, the direct approximation of $\bar{w}([\rho];r)$ is expected to be facilitated by the fact that the change in its exchange-correlation component is continuous at any $r$.

For implications with respect to the band-gap problem in molecules and solids, it is well known [10,11] that $\Delta_{xc}$ is the correction to orbital energy differences for the value of the gap, or ionization energy minus electron affinity, where $\delta$ is, for instance, a fraction of an electron from the valence band that is added to the conduction band in a semiconductor. Then since $w([\rho];r)=\bar{w}([\rho];r)-\lim_{|r|\to\infty}\bar{w}([\rho];r)$ and since $\bar{w}([\rho];r)$ is electron-number continuous, in the sense of a pointwise convergence of a sequence of functions, it follows that the expression for the value of the gap $E_g$ becomes

$$E_g = \bar{\varepsilon}_{N+1}(N)-\bar{\varepsilon}_N(N)+\Delta_{xc}, \qquad (15a)$$

where $\bar{\varepsilon}_N(N)$ and $\bar{\varepsilon}_{N+1}(N)$ are the orbital energies from the N-electron calculation and

$$\begin{aligned}\Delta_{xc} &= \lim_{\delta\to+0}\lim_{|r|\to\infty}\left[\bar{w}([\rho_{N-\delta}];r)-\bar{w}([\rho_{N+\delta}];r)\right]\\ &= \lim_{\delta\to+0}\int\left[\left.\frac{\delta\bar{w}([\rho];r')}{\delta\rho(r)}\right|_{\rho=\rho_{N+\delta}}-\left.\frac{\delta\bar{w}([\rho];r')}{\delta\rho(r)}\right|_{\rho=\rho_{N-\delta}}\right]\rho_N(r')dr'\end{aligned} \qquad (15b)$$

Hence, Eq. (15) provides an alternative way to calculate $E_g$.

[The electron number continuity of $G[\rho]$ follows from the ensemble approach for fractional electron number [6,8]. That is, $G[\rho_{N+\delta}]$ is the convex sum of an N-electron term and an (N+1)-electron term with, respectively, prefactors $(1-\delta)$ and $\delta$. As $\delta\to 0$, the term involving $\delta$ approaches zero, while the term involving $(1-\delta)$ approaches $G[\rho_N]$. The potential $\bar{w}([\rho];r)$ does jump when $\delta$ is very small but finite. However, this jump takes place in very low density regions that have negligible effect upon $\int\bar{w}([\rho];r)\rho(r)dr$.]

Components of $\bar{w}([\rho];r)$ share characteristics of the whole $\bar{w}([\rho];r)$. For instance, let's consider the exchange-correlation energy, $E_{xc}[\rho]$. Then analogous to the relationship of $\bar{w}([\rho];r)$ to $G[\rho]$, one has

$$E_{xc}[\rho] = \int \bar{v}_{xc}([\rho];r)\rho(r)dr, \qquad (16)$$

where

$$\bar{v}_{xc}([\rho];r) = v_{xc}([\rho];r) + \frac{E_{xc}[\rho] - \int v_{xc}([\rho];r)\rho(r)dr}{\int \rho(r)dr} \qquad (17)$$

and where $v_{xc}([\rho];r) = \dfrac{\delta E_{xc}[\rho]}{\delta \rho(r)}$.

The direct-energy KS formulation is readily applicable to spin-density functional theory [12], where the up-spin $\bar{w}_\uparrow([\rho_\uparrow,\rho_\downarrow];r)$ is

$$\bar{w}_\uparrow([\rho_\uparrow,\rho_\downarrow];r) = w_\uparrow([\rho_\uparrow,\rho_\downarrow];r) + \frac{G[\rho_\uparrow,\rho_\downarrow]}{\int(\rho_\uparrow(r)+\rho_\downarrow(r))dr} - \frac{\int w_\uparrow([\rho_\uparrow,\rho_\downarrow];r)\rho_\uparrow(r)dr}{\int \rho_\uparrow(r)dr} \qquad (18)$$

and the down-spin $\bar{w}_\downarrow([\rho_\uparrow,\rho_\downarrow];r)$ is formed by interchanging $\uparrow$ with $\downarrow$ in the above expression.

Several points should be noted about the $\bar{\varepsilon}_i$'s. While the sum of the occupied $\bar{\varepsilon}_i$'s yields nicely the exact interacting ground-state energy of interest, we also have

$$\bar{\varepsilon}_j - \bar{\varepsilon}_i = \varepsilon_j - \varepsilon_i, \qquad (19)$$

so that an orbital energy difference as an approximation to an excitation energy is as valid with the $\bar{\varepsilon}_i$ as with the $\varepsilon_i$. Also, observe that the ionization energy theorem gives

$$I_N = -\bar{\varepsilon}_N + \lim_{r \to \infty} \bar{w}([\rho_N];r), \qquad (20)$$

where $I_N$ and $\rho_N(r)$ are the ionization energy and ground-state density of the N-electron system, which means that

$$\lim_{r \to \infty} \bar{w}\big([\rho_N]; r\big) = \sum_{i=1}^{N-1}\big[\bar{\varepsilon}_i(N-1) - \bar{\varepsilon}_i(N)\big], \qquad (21)$$

where $\bar{\varepsilon}_i(M)$ is the i-th orbital energy of the M-electron system. Equation (21) serves as a valuable constraint for the approximation of $\bar{w}([\rho]; r)$. [See Ref. [13] for examples of constraints in the spirit of Eqs. (20) and (21) in obtaining functional approximations.]

An important practical test for the approximation of $\bar{w}([\rho]; r)$ is the requirement that the value of $\sum_{i=1}^{N} \bar{\varepsilon}_i$ from an output density must not be higher than the value from the corresponding input density, at each direct-energy KS iteration towards self-consistency. This constraint, which embodies several exact properties of $\bar{w}([\rho]; r)$, is dictated by the recent result in Ref. [14].

Since $E_{GS} = \sum_{i=1}^{N} \bar{\varepsilon}_i$, it is worthwhile to employ the $\bar{\varepsilon}_i$'s to asses the quality of a given approximation to $\bar{w}([\rho]; r)$. That is, our recommendation is to target the individual $\bar{\varepsilon}_i$'s for improving functional approximations. $E_{GS}$ will be approximated accurately if the $\bar{\varepsilon}_i$'s are approximated accurately, and the $\bar{\varepsilon}_i$'s provide basic information not contained in just the value of $E_{GS}$. For instance, Table I illustrates that the extent of the error of an approximate $E_{GS}$ alone does not necessarily predict the extent of the errors in the $\bar{\varepsilon}_i$'s when an approximate $\bar{w}([\rho]; r)$ is used. There are surprises in the table.

The values of the $\bar{\varepsilon}_i$'s in Table I are generally, but not always, more accurate than the values of the $\varepsilon_i$'s. What is particularly surprising is the extent of accuracy, with most functionals, of $\bar{\varepsilon}_i$ for the highest-occupied orbital, as exhibited for the Ne atom in Table I. An explanation may provide valuable insight for understanding and improving the functionals.

The infinite solid is typically modeled by a finite unit cell with Bloch-type periodic boundary conditions governed by a three-dimensional $k$-vector of the reciprocal space. If the traditional KS system is replaced with the direct-energy KS system, the eigenvalues become $\bar{\varepsilon}_i(k)$ instead of $\varepsilon_i(k)$. Since only energy differences of $\varepsilon_i(k)$'s can correlate with experimentally observable quantities, the use of $\bar{\varepsilon}_i(k)$ is equivalent to $\varepsilon_i(k)$ in modeling the band structure, while at the same time the $\boldsymbol{k}$-vector-dependent ground-state energy simply becomes $E_{GS}(\boldsymbol{k}) = \sum_{i=1}^{N} \bar{\varepsilon}_i(\boldsymbol{k})$.

Time-dependent DFT (TDDFT) [17] uses orbital energies in terms of pairwise differences as given in Eq. (19). TDDFT also uses the Hartree-exchange-correlation (Hxc) kernel $\frac{\delta^2 G[\rho]}{\delta\rho(\boldsymbol{r})\delta\rho(\boldsymbol{r}')} = \frac{\delta w([\rho];\boldsymbol{r})}{\delta\rho(\boldsymbol{r}')}$. In the direct-energy approach the Hxc-kernel is expressed in terms of $\bar{w}([\rho];\boldsymbol{r})$. Accordingly, take a functional derivative of $w([\rho];\boldsymbol{r}) = \bar{w}([\rho];\boldsymbol{r}) + \int \frac{\delta \bar{w}([\rho];\boldsymbol{r}')}{\delta\rho(\boldsymbol{r})} \rho(\boldsymbol{r}') d\boldsymbol{r}'$ to arrive at

$$\frac{\delta w([\rho];\boldsymbol{r})}{\delta\rho(\boldsymbol{r}')} = \frac{\delta \bar{w}([\rho];\boldsymbol{r})}{\delta\rho(\boldsymbol{r}')} + \int \frac{\delta^2 \bar{w}([\rho];\boldsymbol{r}'')}{\delta\rho(\boldsymbol{r})\delta\rho(\boldsymbol{r}')} \rho(\boldsymbol{r}'') d\boldsymbol{r}'' + \frac{\delta \bar{w}([\rho];\boldsymbol{r}')}{\delta\rho(\boldsymbol{r})}. \qquad (22)$$

Although the first functional derivative of $w([\rho];\boldsymbol{r})$ requires the second functional derivative of $\bar{w}([\rho];\boldsymbol{r})$, the right-hand side of Eq. (22) as opposed to its left-hand side shows the $\boldsymbol{r} \leftrightarrow \boldsymbol{r}'$ symmetry characteristic of the second functional derivative $\frac{\delta^2 G[\rho]}{\delta\rho(\boldsymbol{r})\delta\rho(\boldsymbol{r}')}$. The explicit $\boldsymbol{r} \leftrightarrow \boldsymbol{r}'$ symmetry of the exchange-correlation kernel in terms of $\bar{w}([\rho];\boldsymbol{r})$ is a possible advantage for using an approximate $\bar{w}([\rho];\boldsymbol{r})$ in TDDFT.

The higher functional derivatives of $w([\rho];\boldsymbol{r})$ in terms of $\bar{w}([\rho];\boldsymbol{r})$ for use in TDDFT response theory (for example, TDDFT nuclear gradients and nonlinear optics by TDDFT [~~17~~18]) can be obtained in a similar manner.

In summary, we have observed that each unknown functional in density functional theory may be obtained, in principle, by simply integrating the product of an appropriate potential with the density. For approximation purposes, it is noteworthy that these potentials change gently as the density changes, even when the electron number increases from an integer. As in traditional Kohn-Sham theory, the direct-energy Kohn-Sham formulation generates the exact ground-state energy and density.

The objective is to approximate the whole $\bar{w}([\rho];r)$ directly, rather than its parts, $w([\rho];r)$ and $c[\rho]$, separately. It is the whole $\bar{w}([\rho];r)$ that changes most gently with iso-electronic processes and also does not exhibit a discontinuity. Indeed, as $\delta$ is increased from zero to a small but finite value, the only significant relative change in $\rho_{GS}(r)$ is in its tail, at distant $r$. Since this is the only place where $\bar{w}([\rho];r)$ changes by a relatively constant shift, it means that this change in $\bar{w}([\rho];r)$ is essentially local, which is convenient for approximating $\bar{w}([\rho];r)$. In contrast, the change in $w([\rho];r)$ is nonlocal in that it occurs far from where the relative change in $\rho_{GS}(r)$ is greatest.

One of the first models studied in quantum mechanics is that of non-interacting electrons trapped in a box, where the total ground-state energy is given simply by the sum of orbital energies, with the orbitals occupied according to the Pauli exclusion principle. In this Letter, we have observed that ground-state energies of real physical interacting systems can also be obtained as simple sums of orbital energies. This is accomplished by adding a meaningful universal density-dependent constant to the Kohn-Sham Euler equations. The resultant effective one-body potential, $\bar{w}([\rho];r)$, which acts as a functional derivative in Eq. (6) for energy minimization and density optimization, has several especially desirable features for its approximation. Also, the use of $\bar{w}([\rho];r)$ in Eq. (4) readily allows for the incorporation into $G[\rho]$ of implicit ingredients that depend on the density in an implicit manner, such as orbitals and orbital energies.

We close by emphasizing that it should be clear that there exists a meaningful potential analogous to $\bar{w}([\rho];r)$ that allows one to obtain the exact ground-state density and energy directly from the set of Euler equations of virtually any energy minimization formulation. For example, this would be accomplished by augmenting the Fock-operator in Hartree-Fock theory or by augmenting the single Euler equation for the square-root of the density [19] in orbital-free theories.

| Ne | $\varepsilon_{1s}$ | $\varepsilon_{2s}$ | $\varepsilon_{2p_m}$ $(m=x,y,z)$ |
|---|---|---|---|
| Exact | -30.82 | -1.654 | -0.797 |
| SVWN | -30.30 (+0.52/1.69) | -1.323 (+0.331/20.01) | -0.498 (+0.299/37.52) |
| BLYP | -30.52 (+0.30/0.97) | -1.329 (+0.325/19.65) | -0.491 (+0.306/38.39) |
| PBE | -30.49 (+0.33/1.07) | -1.333 (+0.321/19.41) | -0.490 (+0.307/38.52) |
| M11-L | -31.26 (-0.44/1.43) | -1.555 (+0.099/5.99) | -0.533 (+0.264/33.12) |
| revTPSS | -30.69 (+0.13/0.42) | -1.368 (+0.286/17.29) | -0.504 (+0.293/36.76) |
|  | $\bar{\varepsilon}_{1s}$ | $\bar{\varepsilon}_{2s}$ | $\bar{\varepsilon}_{2p_m}$ |

| Ne | | | $(m=x,y,z)$ | $E_{GS} = \sum_{i=1s,2s,2p_x,2p_y,2p_z} \bar{\varepsilon}_i$ |
|---|---|---|---|---|
| Exact | -36.74 | -7.573 | -6.716 | -128.93 |
| SVWN | -36.50 (+0.24/0.65) | -7.522 (+0.051/0.67) | -6.697 (+0.019/0.28) | -128.23 (+0.70/0.54) |
| BLYP | -36.76 (-0.02/0.05) | -7.561 (+0.012/0.16) | -6.723 (-0.007/0.10) | -128.97 (-0.04/0.03) |
| PBE | -36.72 (+0.02/0.05) | -7.561 (+0.012/0.16) | -6.718 (-0.002/0.03) | -128.87 (+0.06/0.05) |
| M11-L | -37.27 (-0.53/1.44) | -7.568 (+0.005/0.07) | -6.546 (+0.170/2.53) | -128.95 (-0.02/0.02) |
| revTPSS | -36.88 (-0.14/0.38) | -7.558 (+0.015/0.20) | -6.694 (+0.022/0.33) | -129.04 (-0.11/0.08) |

Table I. The exact and approximate values, in hartrees, of $\varepsilon_i$, $\bar{\varepsilon}_i$ and $E_{GS} = \sum_{i=1s,2s,2p_x,2p_y,2p_z} \bar{\varepsilon}_i$ for the Ne atom. The approximate values are computed with a few representative pure (non-hybrid) DFT approximations to $w([\rho];r)$ and to $\bar{w}([\rho];r)$ by means of Eq. (5): LDA (SVWN), GGA (BLYP, PBE) and meta-GGA (M11-L, revTPSS). [The quantum-chemistry computer package General Atomic and Molecular Electronic Structure System (GAMESS) [15] was used for the calculations.] The data from Ref. [16] were adopted for the exact values. The aug-cc-pV6Z basis set was used in the approximate calculations. The error of the approximate value with respect to the exact one is put in parenthesis in the format, "error, relative error magnitude in percent".

The authors thank the National Science Foundation for support from the SI2 Grant No. CHEM -1047772.